\documentclass[a4paper,10pt,twocolumn]{article}
\usepackage[utf8]{inputenc}
\usepackage{url}
\usepackage{graphicx}

\title{An unexpected new explanation of seasonality in suicide attempts: Grey’s Anatomy broadcasting}
\author{Luca Perri$^{1,3}$, Om S. Salafia$^{2,3}$\\ 
\footnotesize{$^{1}$Universit\`a degli Studi dell'Insubria, Via Valleggio, 11, I-22100 Como, Italy}\\
\footnotesize{$^{2}$Universit\`a degli Studi di Milano-Bicocca, Piazza della Scienza 3, I-20126 Milano, Italy}\\
\footnotesize{$^{3}$INAF - Osservatorio Astronomico di Brera Merate, via E. Bianchi 46, I–23807 Merate, Italy}\\}
\date{April 1, 2016}

\begin{document}

\twocolumn[
  \begin{@twocolumnfalse}
    \maketitle
    \begin{abstract}
     Seasonality is one of the oldest and most elucidation-resistant issues in suicide epidemiological research. Despite winter depression (also known as Seasonal Affective Disorder, SAD) is known and treated since many years, worldwide cross-sectional data from 28 countries show a lower frequency of suicide attempts around the equinoxes and a higher frequency in spring (both in Northern and Southern Hemisphere). This peak is not compatible with the SAD explanation. However, in recent years epidemiological research has yielded new results, which provide new perspectives on the matter. In fact, the discovery of a new pathology called Post-Series Depression (PSD) could provide an explanation of the suicide attempts pattern. The aim of this study is to analyse weekly data in order to compare them with the TV series broadcasting. Since medical observations in our sample are distributed over many years, in order to compare them as best as we can with the television programming, Grey's Anatomy series was chosen. This medical drama has been in the top 10 of most viewed TV series since 12 years and it is broadcast all over the world, so that it can be considered a universal and homogeneous phenomenon. A full season of the series is split into two separate units with a hiatus around the end of the calendar year, and it runs from September through May. Data analysis was made in order to prove the correlation between PSD and the increase of suicide attempts. Surprisingly, the data analysis shows that the increase of rate of suicide attempts does not coincide with the breaks in Grey’s Anatomy scheduling, but with the series broadcasting. This therefore suggests that it is the series itself to increase the viewer’s depression.

    \end{abstract}
    
    \vspace{40pt}
    
  \end{@twocolumnfalse}
]

\section{Introduction}
The World Health Organization estimates that 11.4 every 100.000 people (15.0 for males and 8.0 for females) committed suicide around the world in 2012 [1]. For every suicide there are many more people who unsuccessfully attempt suicide every day. It is known that several risk factors act cumulatively to increase vulnerability to suicidal behaviour. Some of them are associated with health system and society, alongside many other reported risk factors.

Despite the complexity of factors that contribute to such behaviour, a seasonal pattern has been described [2]. Indeed, seasonality in suicide rates is a well-known phenomenon and an important topic in epidemiological studies. Interestingly, in a worldwide cross-sectional data from 28 countries, a higher frequency of suicides in spring was found compared to other seasons [2]. Seasonal variation does not apply only to completed suicides. A similar pattern was also reported in suicide attempt. However, some studies report suicide attempt peaks in autumn and/or winter. We performed a search for observational epidemiological studies about seasonality in suicide attempts in PubMed, WebofScience, LILACS and CochraneLibrary databases. Weekly data available were evaluated by rhythmic analysis software. A decrease in suicide attempts in correspondence of the equinoxes was found [2].

The research question of this study is: what causes a higher rate of suicide attempts in spring and in autumn/winter, and a lower rate by the equinoxes? The identification of a global seasonal profile in suicide attempts would provide knowledge to guide governments and public health organizations to develop strategies that can prevent suicide more effectively.

\section{An arduous astronomical explanation}
The tilt of the Earth's axis relative to its orbital plane plays a big role in the weather. The Earth is tilted at an angle of approximately 23.44$^\circ$ to the plane of its orbit, and this causes different latitudes to directly face the Sun as the Earth moves through its orbit. It is this variation that primarily brings about seasons. When Earth is at aphelion (the point in the orbit where our planet is nearest to the Sun) it is winter in the Northern Hemisphere, while the Southern Hemisphere faces the Sun more directly and thus experiences warmer temperatures. Conversely, winter in the Southern Hemisphere occurs when the Earth is at perihelion (the point in the orbit where our planet is farthest from the Sun), and the Northern Hemisphere is better exposed to sunlight. From the perspective of an observer on Earth, the winter Sun has a lower apparent maximum altitude in the sky (i.e.~zenith) than the summer Sun. During winter, in either hemisphere the lower zenith of the Sun causes the sunlight to hit that hemisphere at an oblique angle. In regions experiencing winter, the same amount of solar radiation is spread out over a larger area [3].

Seasonal Affective Disorder (SAD) is a type of depression whose symptoms vary in a seasonal pattern. SAD is sometimes known as "winter depression" because symptoms tend to be more severe during the winter. Symptoms often begin in the autumn as days start getting shorter. They're typically most severe during January and February. SAD symptoms often improve or even disappear in the spring and summer, although they usually return each autumn and winter in a repetitive pattern. The exact cause of SAD is not fully understood, but it is often linked to reduced exposure to sunlight during the shorter autumn and winter days. The most accepted explanation is that a lack of sunlight might prevent the hypothalamus from working properly, which may affect the production of melatonin (a hormone that makes us feel sleepy), the production of serotonin (a hormone that affects our mood, appetite and sleep), and the body's internal clock (circadian rhythm) [4]. 

This could explain the higher rate of suicide attempts in autumn/winter and the lower one by the equinoxes, when day and night hours are approximately equal. But SAD cannot explain the spring peak in suicide attempts, when the duration of the daytime is increasing. For this reason, we seek a different interpretation that could integrate the explanation of the seasonal pattern.

\section{The TV series}
The Post-Series Depression, also known as PSD, is the sadness felt after watching a long series. The bitter feeling when you know the journey is over, but you do not want it to end. This can apply to any series, e.g.~TV series, cartoon series, or even movie series. The effect can also be felt after completing a stand alone piece that is not necessarily part of a series, although this is not as common as PSD is presumably linked to developed attachment to the story characters. Effects include, but are not limited to: a state of depression or sadness, the inability to start another story, the need to re-watch the entire series, Internet abuse linked to the series, creating fan fiction [5].

In North American television, a series is a connected set of television program episodes that run under the same title, possibly spanning many seasons. Since the late 1960s, this broadcast programming schedule typically includes between 20 and 26 episodes. Most new programs for the broadcast networks debut in the ``Fall Season''. Since the 1980s, the ``Fall Season'' normally extends to May. Thus, a ``Full Season'' on a broadcast network now usually runs from September through May for at least 22 episodes. A full season is sometimes split into two separate units with a hiatus around the end of the calendar year. Since the 1990s, these shorter seasons have also been referred to as ``.5'' or half seasons, where the run of shows between September and December is labelled ``Season X'', and the second run between January and May is labelled ``Season X.5'' [6].

A ``season finale'' is the last show of the season, usually characterized by a dramatically suspenseful and uncertain end, called ``cliffhanger''. A cliffhanger is meant to ensure the audience will return and see how the characters resolve the dilemma. Often, the last episode before the hiatus around the end of the calendar year ends with a cliffhanger as well.
We then wondered how these difficult dilemmas or shocking revelations can discourage the viewers, affecting the rate of suicide attempts. Since medical observations in our sample were distributed over many years, in order to compare as best as we can the television programming, the TV series ``Grey's Anatomy'' was chosen. Since 12 years this medical drama is firmly in the top 10 of most viewed TV series and it is broadcast all over the world [7], so that it can be considered a universal and homogeneous phenomenon.

\section{Data analysis}
By averaging the broadcasting dates of the last 12 years, we have established that the average season of Grey's Anatomy is transmitted starting from the 38th week of the year, around the autumn equinox in the Northern Hemisphere, until the winter break of the 49th week. The second half of the season starts from the 2nd week of the new year and continues until the cliffhanger of the 21st week.

We therefore expect an increase in the number of suicide attempts between the 21st and 38th week of the year and around the turn of the new year. The comparison with the data, however, totally denies our theory based on the cliffhanger, suggesting a completely opposite new theory.

The increase in the suicide attempt rate in fact does not coincide with the breaks in Grey’s Anatomy scheduling, but with the series broadcasting. This therefore suggests that it is the series itself, with its troubled love affairs and tense relationships, to increase the viewer’s depression. On the contrary, the season finale is a liberating moment for the viewers, whose suicide attempts dramatically decrease.

\begin{figure}
\begin{center}
 \includegraphics[width=\columnwidth]{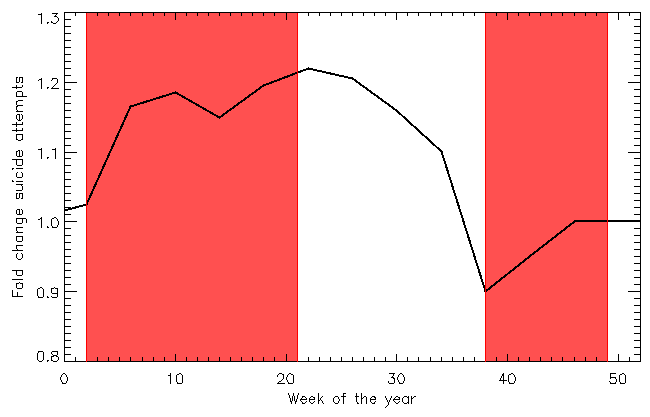}
\end{center}
\caption{\footnotesize{Observed (black line) distribution of suicide attempts (both men and women) in the analyzed sample [2]. The time intervals marked by the red colour are those in which the series is broadcast.}}
 
\end{figure}

As shown in Figure 1, there is a decrease of suicide attempts around the 12th - 13th week of the year, that is around the vernal equinox (but only in the Northern Hemisphere). This peak of optimism related to the arrival of spring and to love in the air, however, is quickly swept away by the first episode of the medical drama, full of destroyed love affairs. The lower rate of suicide attempts around the 38th week of the year is then to be related not so much to the arrival of spring in the Southern Hemisphere as to the absence of Grey’s Anatomy. The rate of suicide attempts during the hiatus around the end of the year is nearly constant: the minor depression related to Grey’s Anatomy is probably balanced by winter depression and sadness related to the Christmas holiday weight gain.

\section{Conclusions}
Our new and unexpected Grey’s Anatomy theory is firmly supported by medical data collected over the last decade. The theory perfectly explains the seasonal pattern in suicide attempts. The observed highest frequency in spring coincides with the season finale. The rate decrease around the equinoxes is also easily explained within the theory.
Future studies could analyse two pending issues: if the approximate constancy of the number of viewers is due to either new addicted people balancing those who committed suicide, or to the poor efficiency of suicide attempts; another interesting question is the actual impact of Derek Shepherd's death on the viewers' depression.

This study, along with others (e.g.~the one that found a correlation between the number of people who drowned after falling into a pool and film appearances by Nicolas Cage [8]), could be a warning to scientists to be wary of spurious correlations. A spurious correlation arises, for example, when measurements which depend on the same variable are compared. In this case, the correlation is simply a consequence of the common dependence of the measurements on that variable, rather than an actual correlation between the measurements. In this case, conclusions would be drawn from a correlation which is actually an artifact of the analysis method, rather than an actual fundamental relationship. But this is obviously not the case.

\section{Acknowledgements}
L. P. and O. S. S. wish to thank Francesco Nappo, Francesca Vimercati, Cristian Consonni, Aaron Gaio and Shonda Rhimes.

\end{document}